# Laminar Film Condensation Heat Transfer on a Vertical, Non-Isothermal, Semi-Infinite Plate

Jian-Jun SHU
School of Mechanical & Aerospace Engineering, Nanyang Technological University,
50 Nanyang Avenue, SINGAPORE 639798
Phone: +65 6790 4459, Fax: +65 6791 1859, E-mail: mjjshu@ntu.edu.sg

***Abstract***

*This paper gives similarity transformations for laminar film condensation on a vertical flat plate with variable temperature distribution and finds analytical solutions for arbitrary Prandtl numbers and condensation rates. The work contrasts with Sparrow and Gregg's assertion that wall temperature variation does not permit similarity solutions. To resolve the long-standing debatable issue regarding heat transfer in the non-isothermal case, some useful formulations are obtained, including a significant dependence of varying Prandtl numbers. Results are compared with the available experimental data.*

## 1 INTRODUCTION

A theory of laminar film condensation was first formulated by Nusselt [1, 2] who considered condensation onto an isothermal flat plate maintained at a constant temperature below the saturation temperature of the surrounding quiescent vapour. Expressions for condensate film thickness and heat transfer characteristics were obtained using simple force and heat balance arguments. Effects due to inertia forces, thermal convection and interfacial shear were neglected, as was surface tension and the possible presence of waves on the condensate film surface. Later, authors have subsequently refined the Nusselt's theory to include some of these omissions. The effects of thermal convection were first examined by Bromley [3] and then by Rohsenow [4] each of whom proposed modifications to the latent heat of condensation to be used in assessing heat transfer at the plate. Sparrow and Gregg [5] were the first to recognize the close parallels between natural convection boundary layers and laminar film condensation. Accordingly, they presented a boundary layer treatment for condensation in the presence of an isothermal cold wall which enabled them to incorporate both thermal convection and inertia effects. Similarity solutions to the governing parabolic equations were derived and detailed numerical solutions were obtained for a wide range of Prandtl numbers and condensation rates. Excellent agreement between the exact numerical results and the Rohsenow's correlation was demonstrated at large Prandtl numbers. The greatest departures from the Nusselt's theory were observed at low Prandtl numbers.

In view of the parallels between natural convection and laminar film condensation, it was a natural progression for Sparrow and Gregg [5] to consider the possible existence of similarity solutions under non-isothermal conditions at the plate. They concluded that the families of wall temperature distributions, power-law $T_w(x)=T^*-T_0\,x^a$ or exponential-law $T_w(x)=T^*-T_0\,e^{bx}$, where $T^*, T_0 > 0$ and $a,b$ are constants, did not in fact permit similarity solutions. In particular, this appeared to preclude the existence of such a solution for the standard boundary condition of constant heat flux at the plate, which they had solved successfully in the natural convection setting in [6]. The influence of variations in wall temperature on the laminar film condensation was studied by Nagendra and Tirunarayanan [7], but unfortunately their presented results were questioned by Subrahmaniyam [8]. The work that follows this assertion is re-examined. Similarity solutions for variable wall temperatures $T_w(x) = T^* - D(x)$ are found and the associated flow and heat transfer characteristics are presented.

## 2 PHYSICAL MODEL AND GOVERNING EQUATIONS

The physical model to be examined is illustrated in Figure 1. A semi-infinite flat plate is aligned vertically with its leading edge uppermost. The surrounding ambient is pure quiescent, saturated vapor maintained at a temperature $T^*$. The plate temperature $T_w(x) = T^* - D(x)$. As a result, the lower temperature at the plate condensation occurs and a continuous laminar film flows downwards along the plate. If the flow is assumed to be in a steady state, the governing equations expressing conservation of mass, momentum and energy are, respectively,

$$\frac{\partial u}{\partial x} + \frac{\partial v}{\partial y} = 0, \quad u\frac{\partial u}{\partial x} + v\frac{\partial u}{\partial y} = g\left(\frac{\rho - \rho^*}{\rho}\right) + \frac{\mu}{\rho}\frac{\partial^2 u}{\partial y^2}, \quad u\frac{\partial T}{\partial x} + v\frac{\partial T}{\partial y} = \frac{k}{\rho C_p}\frac{\partial^2 T}{\partial y^2}, \tag{1}$$

where ($u, v$) are the velocity components associated with the increasing coordinates ($x, y$), measured from the leading edge of the plate along and perpendicular to the plate, and $T$ is the temperature within the condensation film.

Figure 1. Physical model and co-ordinate system

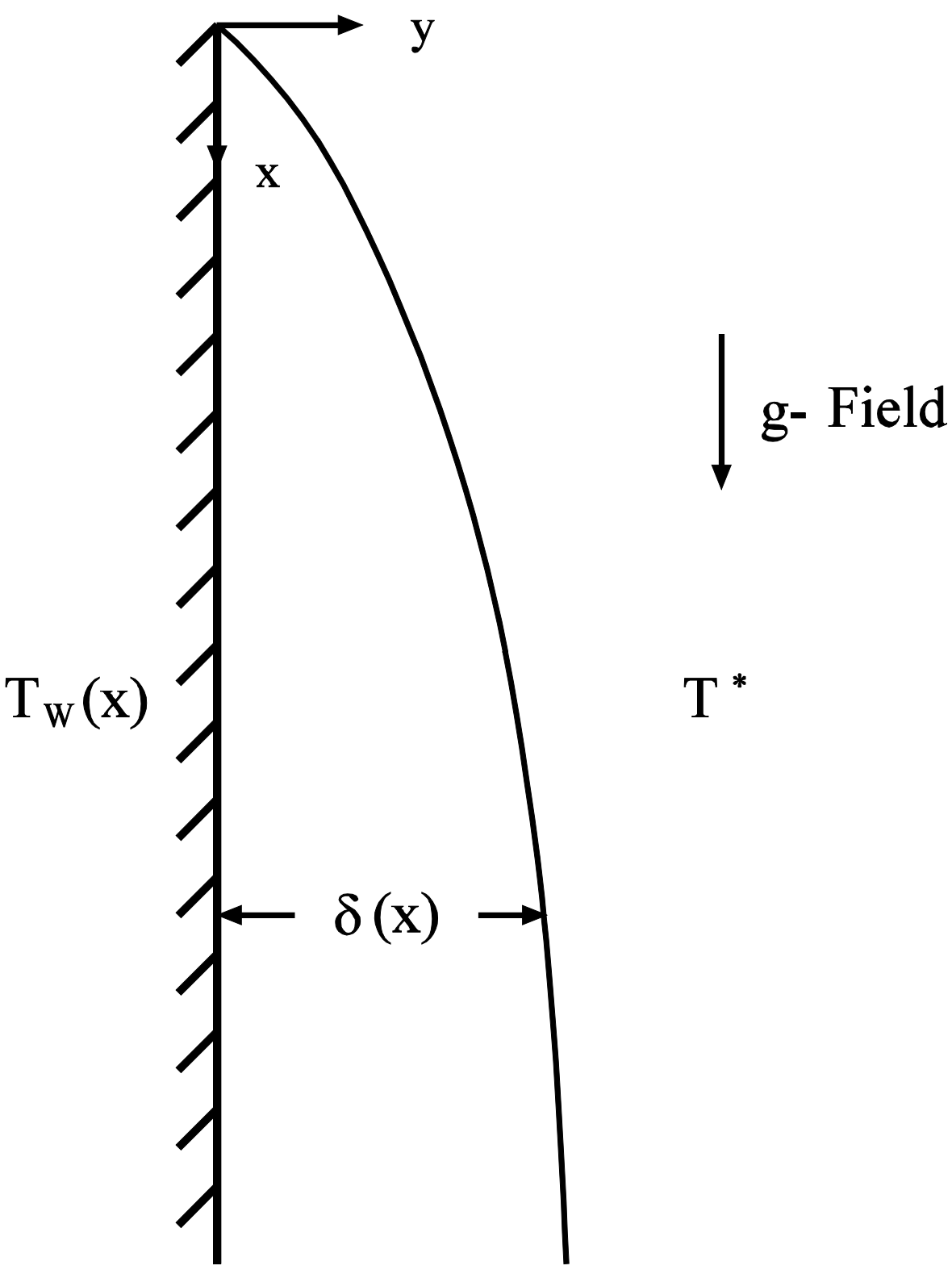

The physical properties, $\rho^*$, the density of the vapor, $\rho, \mu, C_p, k$, the density, the dynamic viscosity, specific heat at constant pressure and the thermal conductivity of the condensate, respectively, are assumed to be constant and $g$ is the acceleration due to gravity. Viscous dissipation and the effects of interfacial shear on the free surface of the condensate are assumed to be negligibly small.

Equations (1) are boundary layer equations derived under the assumption that $\frac{\partial}{\partial x} << \frac{\partial}{\partial y}$ and that the condensation Grashof number,

$$G_{r\,x} = \frac{g\,\rho\left(\rho - \rho^{*}\right)x^{3}}{4\mu^{2}} >> 1.$$

To the level of approximation involved, the $v$-momentum equation is satisfied identically. Flow characteristics are to be established under the hydrodynamic boundary conditions of no slip at the plate and zero shear on the condensate free surface, namely,

$$u = v = 0 \quad \text{on} \quad y = 0; \qquad \frac{\partial u}{\partial y} = 0 \quad \text{on} \quad y = \delta(x), \tag{2}$$

where $\delta(x)$ is the thickness of the condensate film. The prescribed temperature conditions for variable temperatures at the plate and saturation temperature on the free surface require [9, 10]

$$T = T_{w}(x) = T^{*} - D(x) \quad \text{on} \quad y = 0; \qquad T = T^{*} \quad \text{on} \quad y = \delta(x). \tag{3}$$

## 3 SIMILARITY EQUATIONS

In practice, this function $D(x)$ is a holomorphic function and can be expanded into a convergent power series, so assuming

$$D(x) = \sum_{n=0}^{+\infty} T_{n} x^{p_{n}} , \tag{4}$$

where $\{T_{n}\}$ is a sequence of numbers and $\{p_{n}\}$ is a non-negative increasing sequence. The continuity equation in (1) is identically satisfied if a stream function formulation

$$u = \frac{\partial \psi}{\partial y}, \qquad v = -\frac{\partial \psi}{\partial x},$$

is introduced. It then remains to invoke, if possible, similarity transformations which reduce the parabolic, partial differential equations (1) and their boundary conditions to ordinary differential equations. Because the third equation in (1) is linear for temperature $T$, this possibility is readily examined by the preliminary transformations

$$\psi \sim x^{r} f(\eta), \qquad \eta \sim \frac{y}{x^{s}} \quad \text{and} \quad T \sim x^{p_{n}} \theta_{n}(\eta).$$

The second equation in (1) reduces to an ordinary differential equation in $\eta$ only if $2r - 2s - 1 = r - 3s = 0$ *i.e.* $s = \frac{1}{4}$, $r = \frac{3}{4}$. Such results are entirely in keeping with the findings of Sparrow and Gregg [5]. If the third equation in (1) is also to become an ordinary differential equation in $\eta$ then $r + p_{n} - s - 1 = p_{n} - 2s$. Thus, $s = \frac{1}{4}$, $r = \frac{3}{4}$ and $p_{n}$ provide similarity scalings which ensure that the governing equations do indeed reduce to a pair of ordinary differential equations. To examine solutions at the extremes for low and high Prandtl numbers, $P_{r} = \frac{\mu C_{p}}{k}$, two forms of transformation are instructive.

$$P_{r} \leq 1: \quad \psi = \frac{4\mu}{\rho} G_{r\,x}^{\frac{1}{4}} f(\eta) \qquad\qquad P_{r} > 1: \quad \psi = \frac{4\mu}{\rho} \frac{G_{r\,x}^{\frac{1}{4}}}{P_{r}^{\frac{3}{4}}} \tilde{f}(\tilde{\eta}) \tag{5a}$$

$$\eta = \frac{y}{x} G_{rx}^{\frac{1}{4}} \qquad \tilde{\eta} = \frac{y}{x} G_{rx}^{\frac{1}{4}} P_r^{\frac{1}{4}} \tag{5b}$$

$$T - T^* = -\sum_{n=0}^{+\infty} T_n x^{P_n} \theta_n(\eta) \qquad T - T^* = -\sum_{n=0}^{+\infty} T_n x^{P_n} \tilde{\theta}_n(\tilde{\eta}). \tag{5c}$$

Equations (1) reduce to

$$f''' + 3ff'' - 2f'^2 + 1 = 0 \qquad \tilde{f}''' + \frac{1}{P_r}\left(3\tilde{f}\tilde{f}'' - 2\tilde{f}'^2\right) + 1 = 0 \tag{6a}$$

$$\theta_n'' + P_r\left(3f\theta_n' - 4p_n f'\theta_n\right) = 0 \qquad \tilde{\theta}_n'' + 3\tilde{f}\tilde{\theta}_n' - 4p_n\tilde{f}'\tilde{\theta}_n = 0, \tag{6b}$$

with boundary conditions

$$f(0) = f'(0) = 0; \quad \theta_n(0) = 1; \qquad \tilde{f}(0) = \tilde{f}'(0) = 0; \quad \tilde{\theta}_n(0) = 1; \tag{7a}$$

$$f''(\eta_\delta) = \theta_n(\eta_\delta) = 0; \qquad \tilde{f}''(\tilde{\eta}_\delta) = \tilde{\theta}_n(\tilde{\eta}_\delta) = 0, \tag{7b}$$

where

$$\tilde{f} = P_r^{\frac{3}{4}} f, \qquad \tilde{\eta} = P_r^{\frac{1}{4}} \eta, \qquad \tilde{\theta}_n = \theta_n. \tag{8}$$

$\eta_\delta$ denotes the value of $\eta$ on the condensate free surface $y = \delta(x)$ and ' here denotes $\dfrac{d}{d\tilde{\eta}}$.

$$u = \frac{4\mu G_{rx}^{\frac{1}{2}}}{\rho x}\frac{\partial f}{\partial \eta} = \frac{4\mu G_{rx}^{\frac{1}{2}}}{\rho P_r^{\frac{1}{2}} x}\frac{\partial \tilde{f}}{\partial \tilde{\eta}}, \quad v = \frac{\mu G_{rx}^{\frac{1}{4}}}{\rho x}\left[\eta\frac{\partial f}{\partial \eta} - 3f\right] = \frac{\mu G_{rx}^{\frac{1}{4}}}{\rho P_r^{\frac{3}{4}} x}\left[\tilde{\eta}\frac{\partial \tilde{f}}{\partial \tilde{\eta}} - 3\tilde{f}\right]. \tag{9}$$

Let the condensate film flow rate per unit width of the wall be

$$Q_x = \int_0^{\delta(x)} \rho u \, dy. \tag{10}$$

Then, from (5) and (9),

$$Q_x = 4\mu G_{rx}^{\frac{1}{4}} f(\eta_\delta) = 4\mu P_r^{-\frac{3}{4}} G_{rx}^{\frac{1}{4}} \tilde{f}(\tilde{\eta}_\delta). \tag{11}$$

Although equations (6) and (7) may be solved quite satisfactorily for prescribed $P_r$ and $\eta_\delta$ or $\tilde{\eta}_\delta$, it remains to relate the non-dimensional condensate film thicknesses to the prevailing physical conditions under which condensation occurs. Such conditions are reflected by the condensation parameter $\dfrac{C_p \Delta T}{h_{fg}}$ where $C_p$ is the specific heat at constant pressure, $h_{fg}$ is the latent heat of condensation and $\Delta T$ is the local temperature differential between $T^*$ and the temperature at the plate. The required correlation is obtained from the overall energy balance

$$-\int_0^x \kappa\left(\frac{\partial T}{\partial y}\right)_{y=0} dx + \int_0^{\delta(x)} \rho u h_{fg} \, dy + \int_0^{\delta(x)} \rho u C_p \left(T^* - T\right) dy = 0, \tag{12}$$

where the first term is the heat transferred from the condensate to the plate, the second term represents the latent heat of condensation, and the third term is due to the subcooling of the condensate below its saturation temperature. Negligible heat conduction across the liquid-vapour interface has been assumed. In terms of transformed variables, this leads to the result

$$\sum_{n=0}^{+\infty} \frac{C_p \Delta T_n}{h_{fg}} \frac{\theta_n'(\eta_\delta)}{4p_n + 3} = -P_r f(\eta_\delta) \qquad \text{or} \qquad \sum_{n=0}^{+\infty} \frac{C_p \Delta T_n}{h_{fg}} \frac{\tilde{\theta}_n'(\eta_\delta)}{4p_n + 3} = -\tilde{f}(\eta_\delta), \tag{13}$$

where $\Delta T_n = T_n x^{p_n}$, *i.e.*, $T^* - T_w(x) = \sum_{n=0}^{+\infty} \Delta T_n$.

The quantities that appear in equation (13) are output data from the solutions of the appropriate equations. As pointed out by Sparrow and Gregg [5], a unique correlation, therefore, exists between the condensation parameter and $\eta_\delta$ or $\tilde{\eta}_\delta$.

## 4 ANALYTICAL SOLUTIONS

Equations (6) and (7) can be solved analytically rather than numerically through series solutions, so that some useful formulas and properties of condensation heat transfer can be easily found. The asymptotic solutions of the equations (6) and (7) are

$$f = \eta^2 \left[ \frac{1}{2}\eta_\delta - \frac{1}{12}\eta_\delta^5 - \frac{1}{6}\eta + \frac{1}{120}\eta_\delta^2 \eta^3 + O\left(\eta_\delta^9\right) \right] \tag{14}$$

$$\theta_n = 1 - \left[ \frac{1}{\eta_\delta} + \frac{P_r}{30}\left(3 + 8p_n\right)\eta_\delta^3 \right]\eta + \frac{2P_r}{3} p_n \eta_\delta \eta^3 + \frac{P_r}{8}\left(1 - 4p_n\right)\eta^4 - \frac{P_r}{40}\left(1 - 4p_n\right)\frac{\eta^5}{\eta_\delta} + O\left(\eta_\delta^8\right) \tag{15}$$

$$\tilde{f} = \tilde{\eta}^2 \left[ \frac{1}{2}\tilde{\eta}_\delta - \frac{1}{12P_r}\tilde{\eta}_\delta^5 - \frac{1}{6}\tilde{\eta} + \frac{1}{120P_r}\tilde{\eta}_\delta^2 \tilde{\eta}^3 + O\left(\tilde{\eta}_\delta^9\right) \right] \tag{16}$$

$$\tilde{\theta}_n = 1 - \left[ \frac{1}{\tilde{\eta}_\delta} + \frac{1}{30}\left(3 + 8p_n\right)\tilde{\eta}_\delta^3 \right]\tilde{\eta} + \frac{2}{3} p_n \tilde{\eta}_\delta \tilde{\eta}^3 + \frac{1}{8}\left(1 - 4p_n\right)\tilde{\eta}^4 - \frac{1}{40}\left(1 - 4p_n\right)\frac{\tilde{\eta}^5}{\tilde{\eta}_\delta} + O\left(\tilde{\eta}_\delta^8\right). \tag{17}$$

## 5 APPROXIMATE FORMULAS

The most important result given in this present type of problem is the expression for heat transfer. This can be most conveniently presented by introducing a local Nusselt number

$$N_{ux} = \frac{x}{T^* - T_w(x)} \left.\frac{\partial T}{\partial y}\right|_{y=0}, \tag{18}$$

where $T_w(x)$ is the wall temperature. The Nusselt number and the Grashof number are traditionally combined in one expression, which in our case leads to

$$\frac{N_{ux}}{G_{rx}^{\frac{1}{4}}} = -\frac{\sum_{n=0}^{+\infty} \frac{C_p \, \Delta T_n}{h_{fg}} \theta_n'(0)}{\sum_{n=0}^{+\infty} \frac{C_p \, \Delta T_n}{h_{fg}}} \quad \text{or} \quad \frac{N_{ux}}{G_{rx}^{\frac{1}{4}}} = -\frac{P_r^{\frac{1}{4}} \sum_{n=0}^{+\infty} \frac{C_p \, \Delta T_n}{h_{fg}} \tilde{\theta}_n'(0)}{\sum_{n=0}^{+\infty} \frac{C_p \, \Delta T_n}{h_{fg}}}. \tag{19}$$

The asymptotic approximation is

$$\frac{N_{ux}}{G_{rx}^{\frac{1}{4}}} \approx \left( \sum_{n=1}^{+\infty} \frac{3}{3 + 4p_n} \frac{C_p \, \Delta T_n}{P_r \, h_{fg}} \right)^{-\frac{1}{4}}, \tag{20}$$

for arbitrary wall temperature variations when $P_r \to 0$ or $P_r \to +\infty$.

## 6 POWER-LAW TEMPERATURE VARIATION

For the case of the power-law temperature variation, we have $D(x) = T_0 \, x^a$, where $T_0 > 0$ and $a$ are two constants, so that (13) and (19) give

$$\frac{C_p \Delta T}{h_{fg}} = -(4a+3) P_r \frac{f(\eta_\delta)}{\theta'(\eta_\delta)} = -(4a+3) \frac{\tilde{f}(\eta_\delta)}{\tilde{\theta}'(\eta_\delta)}, \tag{21}$$

where $\Delta T = T^* - T_w(x) = T_0\, x^a$ and

$$\frac{N_{u\,x}}{G_{r\,x}^{\frac{1}{4}}} = -\theta'(0) = -P_r^{\frac{1}{4}}\, \tilde{\theta}'(0). \tag{22}$$

Hence it can be shown that

$$N_{u\,x} \left[ \frac{\dfrac{C_p\, \Delta T}{h_{fg}}}{\dfrac{g\, C_p\, \rho(\rho-\rho^*) x^3}{4\,\mu\,\kappa}} \right]^{\frac{1}{4}} \approx \left( \frac{3+4a}{3} \right)^{\frac{1}{4}} \left\{ 1 - \frac{3\left[ 9 - (27+52a) P_r \right]}{160(3+4a)} \frac{C_p\, \Delta T}{P_r\, h_{fg}} \right\}, \tag{23}$$

for low or high Prandtl numbers. It is worth to mention to this end that the corrected power-law Nusselt's formula concerning the questionable results presented in [7, 8] should be

$$\frac{N_{u\,x}(a)}{N_{u\,x}(a=0)} \approx \left( \frac{3+4a}{3} \right)^{\frac{1}{4}}. \tag{24}$$

Most of the existing experimental data deal with the average heat transfer coefficient. It is, therefore, necessary for the sake of comparison between theory and experiment to consider the definition of the average heat transfer coefficient for the case where the surface temperature varies as in the present case.

Since $\left( \dfrac{\mu^2}{g\,\rho^2} \right)^{\frac{1}{3}}$ has a unit of length, an average Nusselt number may be defined by

$$N_{u\,m}^* = \frac{q_m}{\kappa\, \Delta T_m} \left( \frac{\mu^2}{g\,\rho^2} \right)^{\frac{1}{3}}, \tag{25}$$

and the Reynolds number may be defined by

$$R_e^* = \frac{4}{\mu} \frac{q_m\, L}{h_{fg}}, \tag{26}$$

where the mean temperature difference is

$$\Delta T_m = \frac{1}{L} \int_0^L \Delta T\, dx, \tag{27}$$

and the average heat transfer rate is

$$q_m = \frac{1}{L} \int_0^L \left( \kappa \frac{\partial T}{\partial y} \bigg|_{y=0} \right) dx. \tag{28}$$

For $R_e^* < \dfrac{16}{3} P_r^{-\frac{3}{4}}\, G_{r\,L}^{\frac{1}{4}}$, the asymptotic approximation becomes

$$N_{u\,m}^* \approx \frac{4}{3} \left( \frac{\rho-\rho^*}{\rho} \right)^{\frac{1}{3}} \left( \frac{3}{4} R_e^* \right)^{-\frac{1}{3}} \left[ 1 + \frac{81 + 264\,a + 208\,a^2 - \dfrac{9}{P_r}(3+4a)}{120(3+7a)} \left( \frac{3\, R_e^*\, P_r^{\frac{3}{4}}}{16\, G_{r\,L}^{\frac{1}{4}}} \right)^{\frac{4}{3}} \right]. \tag{29}$$

Letting $\dfrac{3\, R_e^*\, P_r^{\frac{3}{4}}}{16\, G_{r\,L}^{\frac{1}{4}}} << 1$ and $\rho >> \rho^*$, (29) becomes

$$N_{u\,m}^{*} \approx \frac{4}{3}\left(\frac{3}{4}R_e^{*}\right)^{-\frac{1}{3}}, \tag{30}$$

which agrees with Fujii *et al.*'s approximate solution [11] for the case of uniform surface heat flux.

## 7 ISOTHERMAL CASE AND COMPARISONS OF PREVIOUS CORRELATIONS AND EXPERIMENTAL VALUES

The previous theoretical analyses of laminar film condensation mostly focused on the isothermal surface. In the isothermal case, $D(x) = \text{constant}$ such that $p_n = 0$ for any $n$. The asymptotic approximate becomes

$$N_{u\,x}\left[\frac{\dfrac{C_p\,\Delta T}{h_{fg}}}{\dfrac{g\,C_p\,\rho\left(\rho-\rho^{*}\right)x^{3}}{4\,\mu\,\kappa}}\right]^{\frac{1}{4}} = 1 + \frac{9\left(3-\dfrac{1}{P_r}\right)}{160}\frac{C_p\,\Delta T}{h_{fg}} - \frac{39355-\dfrac{9650}{P_r}-\dfrac{7069}{P_r^{\,2}}}{1075200}\left(\frac{C_p\,\Delta T}{h_{fg}}\right)^{2} + O\left[\left(\frac{C_p\,\Delta T}{h_{fg}}\right)^{3}\right]. \tag{31}$$

The comparison between the approximate formula (31) and the exact numerical result is shown in Table 1. These Nusselt number results for large and low Prandtl numbers are presented in Figure 2. For large values of $\dfrac{C_p\,\Delta T}{h_{fg}}$, the approximate formula (31) deviates from the exact numerical result, because the truncation error in (31) is $O\left[\left(\dfrac{C_p\,\Delta T}{h_{fg}}\right)^{3}\right]$. From the coefficients of the first degree and the second degree of $\dfrac{C_p\,\Delta T}{h_{fg}}$ in (31), we know that there are two special values $P_{r1} = \dfrac{1}{3}$ and $P_{r2} \approx 0.56379651$. When $P_r \geq P_{r2}$, the curves are monotone increasing and convex. When $P_{r2} > P_r \geq P_{r1}$, the curves are monotone increasing but concave. When $P_r < P_{r1}$, the curves are monotone decreasing and concave. Comparing with the Nusselt's simple theory in the type of problem [1, 2], which predicted that

$$N_{u\,x}\left[\frac{\dfrac{C_p\,\Delta T}{h_{fg}}}{\dfrac{g\,C_p\,\rho\left(\rho-\rho^{*}\right)x^{3}}{4\,\mu\,\kappa}}\right]^{\frac{1}{4}} = 1, \tag{32}$$

the result shows formula (32) is satisfied only at $P_r = \dfrac{1}{3}$. As the condensate film thickness increases, inertia forces tend to decrease the heat transfer, while sub-cooling tends to increase the heat transfer. When $P_r > \dfrac{1}{3}$, sub-cooling wins out over inertia forces and the curves rise monotonically. When $P_r < \frac{1}{3}$, inertia forces win out over sub-cooling and the curves fall monotonically. When $P_r = \frac{1}{3}$, sub-cooling first matches and then wins out over inertia forces.

Table 1. Comparison of approximate formula with exact numerical result for Prandtl number $= 2.58$

| $\dfrac{C_p \Delta T}{h_{fg}}$ | $N_{u\,x}\left[\dfrac{\dfrac{C_p\,\Delta T}{h_{fg}}}{\dfrac{g\,C_p\,\rho\left(\rho-\rho^*\right)x^3}{4\,\mu\,\kappa}}\right]^{\frac{1}{4}}$ | |
|---|---|---|
| | Exact result | Approximate formula (31) |
| 0.0001 | 1.0000 | 1.0000 |
| 0.0016 | 1.0002 | 1.0002 |
| 0.0081 | 1.0012 | 1.0012 |
| 0.0257 | 1.0038 | 1.0038 |
| 0.0632 | 1.0092 | 1.0092 |
| 0.1328 | 1.0190 | 1.0189 |
| 0.2511 | 1.0350 | 1.0349 |
| 0.4419 | 1.0595 | 1.0587 |
| 0.7402 | 1.0947 | 1.0912 |
| 1.1997 | 1.1431 | 1.1300 |
| 1.9047 | 1.2076 | 1.1633 |
| 2.9923 | 1.2908 | 1.1520 |

Figure 2. Comparison of approximate formula (31) with exact numerical result for various Prandtl numbers $P_r =$ (a) $0.003$ (b) $\dfrac{1}{3}$ (c) $0.45$ (d) $0.56379651$ (e) $1$ (f) $10$ in terms of

$$N_{ux}\left[\frac{\dfrac{C_p\,\Delta T}{h_{fg}}}{\dfrac{g\,C_p\,\rho\left(\rho-\rho^*\right)x^3}{4\,\mu\,\kappa}}\right]^{\frac{1}{4}} \quad \text{versus} \quad \frac{C_p\,\Delta T}{h_{fg}}$$

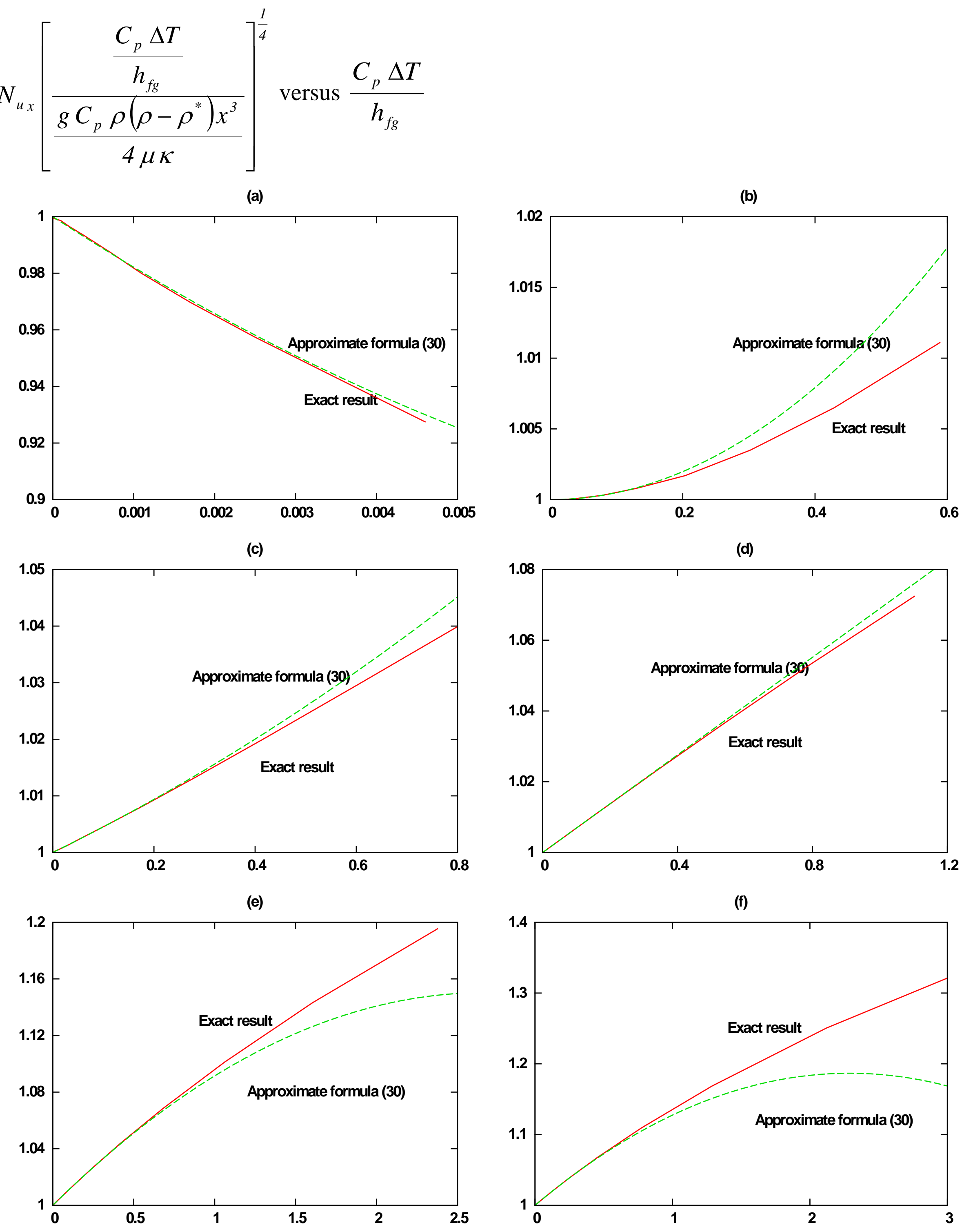

Letting $P_r \to +\infty$, (31) becomes

$$N_{u\,x}\left[\frac{\dfrac{C_p\,\Delta T}{h_{fg}}}{\dfrac{g\,C_p\,\rho\left(\rho-\rho^*\right)x^3}{4\,\mu\,\kappa}}\right]^{\frac{1}{4}}=\left\{1+\frac{27}{40}\frac{C_p\,\Delta T}{h_{fg}}+\frac{1643}{67200}\left(\frac{C_p\,\Delta T}{h_{fg}}\right)^2+O\left[\left(\frac{C_p\,\Delta T}{h_{fg}}\right)^3\right]\right\}^{\frac{1}{4}}$$

$$\approx\left[1+0.675\frac{C_p\,\Delta T}{h_{fg}}+0.024449404\left(\frac{C_p\,\Delta T}{h_{fg}}\right)^2\right]^{\frac{1}{4}}. \tag{33}$$

If the Rohsenow's correlation [4] is now referred, it is apparent that

$$N_{u\,x}\left[\frac{\dfrac{C_p\,\Delta T}{h_{fg}}}{\dfrac{g\,C_p\,\rho\left(\rho-\rho^*\right)x^3}{4\,\mu\,\kappa}}\right]^{\frac{1}{4}}=\left(1+0.68\frac{C_p\,\Delta T}{h_{fg}}\right)^{\frac{1}{4}}, \tag{34}$$

which is identical to this paper's result, when the inertia forces are neglected, that is, the Prandtl number approaches infinity and only the first approximation is considered. It is quite evident that the correlation (31) is more accurate than previous correlations, especially at finite and low Prandtl numbers. For low or high Prandtl numbers and $R_e^* < \frac{16}{3} P_r^{-\frac{3}{4}} G_{r\,L}^{\frac{1}{4}}$, (29) is simplified as

$$N_{u\,m}^* \approx \frac{4}{3}\left(\frac{\rho-\rho^*}{\rho}\right)^{\frac{1}{3}}\left(\frac{3}{4}R_e^*\right)^{-\frac{1}{3}}\left[1+\frac{3\left(3-\dfrac{1}{P_r}\right)}{40}\left(\frac{3\,R_e^*\,P_r^{\frac{3}{4}}}{16\,G_{r\,L}^{\frac{1}{4}}}\right)^{\frac{4}{3}}\right]. \tag{35}$$

Letting $\dfrac{3\,R_e^*\,P_r^{\frac{3}{4}}}{16\,G_{r\,L}^{\frac{1}{4}}} << 1$ and $\rho >> \rho^*$, (35) becomes

$$N_{u\,m}^* = \frac{4}{3}\left(\frac{3}{4}R_e^*\right)^{-\frac{1}{3}}, \tag{36}$$

which is the same as the Nusselt formula [1, 2] for the case of uniform surface temperature. In the Table 2, $L$ expresses an effective length of flat plate or tube. $E$ and $E_N$ express, respectively, the Nusselt number relative errors of formula (35) and the Nusselt formula (36) against experiment. The property values for water and steam are taken from Kaye & Laby [12] and Weast [13]. Table 2 shows that the present formula (35) predicts the experimental results better than the Nusselt's formula (36) except Mills and Seban's case, where their tube was too short. Unfortunately, the above-measured experimental data are confined to $\dfrac{3\,R_e^*\,P_r^{\frac{3}{4}}}{16\,G_{r\,L}^{\frac{1}{4}}} < 0.2$. The improvement of the present formula (35) is not notable.

Table 2. Comparisons of Nusselt numbers of various observers condensing pure saturated steam on vertical flat plates or tubes

| Observers | $L$ | $P_r$ | $R_e^*$ | $G_{rL}$ | $\frac{E}{E_N}$ |
|---|---|---|---|---|---|
| Baker *et al.* [14] | 20 ft | 1.4823 | 510.65 ~ 5734.08 | $8.6565\times10^{15}$ | $\frac{18.10}{18.38}$ |
| Meisenburg *et al.* [15] | 12 ft | 1.5072 | 216.80 ~ 4730.85 | $1.8108\times10^{15}$ | $\frac{26.21}{26.60}$ |
| Hebbard & Badger [16] | 11.969 ft | 1.5274 | 632.61 ~ 3557.93 | $1.7518\times10^{15}$ | $\frac{31.93}{32.26}$ |
| Garrett & Wighton [17] | 12.5 in | 2.1170 | 247.74 ~ 327.50 | $6.4192\times10^{11}$ | $\frac{12.40}{13.20}$ |
| Mills & Seban [18] | 5 in | 6.2625 | 7.18 ~ 8.02 | $6.0926\times10^{9}$ | $\frac{-17.95}{-17.81}$ |

## 8 EXPONENTIAL-LAW TEMPERATURE VARIATION

For the case of exponential-law temperature variation, we have

$$D(x) = T_0\, e^{b\,x} = T_0 \sum_{n=0}^{+\infty} \frac{b^n}{n!} x^n\,,$$

where $T_0 > 0$ and $b$ are two constants. The following results can be obtained.

$$N_{u\,x}\left[\frac{\dfrac{C_p\,\Delta T}{h_{fg}}}{\dfrac{g\,C_p\,\rho\left(\rho-\rho^*\right)x^3}{4\,\mu\,\kappa}}\right]^{\frac{1}{4}} \approx K^{-\frac{1}{4}}\left\{1-\frac{1}{480}\left\{27\,K-\left[21+4\,K\left(15+32\,b\,x\right)\right]P_r\right\}\frac{C_p\,\Delta T}{P_r\,h_{fg}}\right\}, \tag{37}$$

where $K = M\left(1;\dfrac{7}{4};-b\,x\right)$. $M(\alpha;\beta;x)$ denotes the Kummer function and is usually defined as [19]

$$M(\alpha;\beta;x) = \sum_{n=0}^{+\infty}\frac{(\alpha)_n}{n!\,(\beta)_n}x^n\,,$$

and $(x)_n$ denotes the Pochhammer polynomial and is defined as the $n$-fold product [19]

$$(x)_n = x(x+1)(x+2)\cdots(x+n-1) \qquad n = 1,2,\cdots,$$

with $(x)_0 = 1$. It is worth to mention to this end that the corrected exponential-law Nusselt's formula concerning the questionable results presented in [7, 8] should be

$$\frac{N_{u\,x}(b)}{N_{u\,x}(b=0)} \approx \left[\frac{1}{M\left(1;\dfrac{7}{4};-bx\right)}\right]^{\frac{1}{4}}. \tag{38}$$

## 9 CONCLUDING REMARKS

A detailed examination of laminar film condensation against a non-isothermal vertical plate has been performed. Similarity transformations for laminar film condensation on a vertical flat plate and analytical solutions for arbitrary Prandtl numbers and condensation rates have been given here. Based on the similarity solutions, the respective formulations for low and high Prandtl numbers have been followed in parallel and asymptotic formulae for the associated heat transfer characteristics have been obtained. For specific physical conditions, the results contract to the results of earlier workers thus providing a basic confirmation of the validity of the solutions obtained. Unfortunately, relevant experimental information is very scarce and often results have been surmised from inappropriate physical configurations and ambient conditions. Currently, there is no satisfactory agreement between even first-order results and experimental results. Accordingly, it is difficult to gauge the significance and value of higher order terms. Nevertheless, the range of non-isothermal conditions for which heat transfer estimates may be obtained has been extended and the associated asymptotic representations presented. Results are compared with the available experimental data. As a separate, but potentially related and extremely interesting application, the various problems [20-29] are involving the film formation due to condensation on non-isothermal conditions. To resolve the long debatable issue regarding the ratio of heat transfer results in the non-isothermal (NS) case with those in isothermal (IS) case, the corrected Nusselt's formula should be $\frac{N_{u\,\mathrm{NS}}}{N_{u\,\mathrm{IS}}} \approx \left(\frac{3+4a}{3}\right)^{\frac{1}{4}}$ for power-law wall temperature distribution $T_w(x) = T^* - T_0\, x^a$ and $\frac{N_{u\,\mathrm{NS}}}{N_{u\,\mathrm{IS}}} \approx \left[\frac{1}{M\left(1;\frac{7}{4};-bx\right)}\right]^{\frac{1}{4}}$ with the Kummer function $M\left(1;\frac{7}{4};-b\,x\right)$ for exponential-law wall temperature distribution $T_w(x) = T^* - T_0\, e^{b\,x}$, respectively.